\documentclass{amsart}
\usepackage{amsmath,amssymb,amsthm}
\usepackage{graphicx}

\newcommand{\be}{\begin{equation}}
\newcommand{\ee}{\end{equation}}
\newcommand{\bea}{\begin{array}}
\newcommand{\ea}{\end{array}}
\newcommand{\beqa}{\begin{eqnarray}}
\newcommand{\eeqa}{\end{eqnarray}}
\newcommand{\bean}{\begin{eqnarray*}}
\newcommand{\eean}{\end{eqnarray*}}

\def\up#1{\leavevmode \raise.16ex\hbox{#1}}

\newcommand{\gapproxeq}{\lower
 .7ex\hbox{$\;\stackrel{\textstyle >}{\sim}\;$}}
\newcommand{\lapproxeq}{\lower .7ex\hbox{$\;\stackrel
{\textstyle <}{\sim}\;$}}

\newcounter{appendice}

\def\thebibliography#1{{\bf REFERENCES\markboth
 {REFERENCES}{REFERENCES}}\list
 {[\arabic{enumi}]}{\settowidth\labelwidth{[#1]}\leftmargin\labelwidth
 \advance\leftmargin\labelsep
 \usecounter{enumi}}
 \def\newblock{\hskip .11em plus .33em minus -.07em}
 \sloppy
 \sfcode`\.=1000\relax}

\begin{document}
\centerline{\Large On deformed quantum mechanical schemes and $\star$-value equations}
\centerline{\Large
 based on the space-space noncommutative Heisenberg-Weyl group}
\vskip .5cm
\centerline{ {L. Rom\'an Ju\'arez~$^a$} and {Marcos Rosenbaum~$^b$} }
\vskip 1cm
\begin{center}
Instituto de Ciencias Nucleares,\\
Universidad Nacional Aut\'onoma de M\'exico,\\
 A. Postal 70-543 , M\'exico D.F., M\'exico \\
{\it a)roman.juarez@nucleares.unam.mx,\\ 
b)mrosen@nucleares.unam.mx\\
}
\vskip 1cm
\centerline{Published in J. Phys. Math. Dec. 2010}
\end{center}


\vskip .5cm \vspace*{5mm} \normalsize \centerline{\bf ABSTRACT}
\vspace{.5cm}

We investigate the Weyl-Wigner-Gr\"oenewold-Moyal, the Stratonovich and the Berezin  group quantization schemes for the  space-space noncommutative Heisenberg-Weyl group. 
We show that the $\star$-product for the deformed algebra of Weyl functions for the first scheme is different than that for the other two, even though their respective quantum
mechanics' are equivalent as far as expectation values are concerned, provided that some additional criteria are imposed on the implementation of this process. We also show that
it is the $\star$-product associated with the Stratonovich and the Berezin formalisms that correctly gives the Weyl symbol of a product of operators in terms of the deformed 
product of their corresponding Weyl symbols. To conclude, we derive  the stronger $\star$-valued equations for the 3 quantization schemes considered and discuss the criteria 
that are also needed for them to exist.

\vspace{2cm}
\newpage


\section{Introduction}
It is well known \cite{weyl,wigner,gron,moy} that for non-relativistic standard Quantum Mechanics the expectation value of an operator on Hilbert space can be formally represented as
a statistical-like average of the corresponding Weyl phase-space function with the statistical density given by the Wigner function
associated with the density matrix of the quantum state. Moreover, when applying this scheme to a product of two arbitrary operator functions
of the quantum position and momentum operators their corresponding Weyl
 phase-space function was given by the exponential of the Poisson bi-differential acting on the Weyl equivalent of each of the two
operators. This correspondence between the product of quantum operators and the twisted product of their classical phase-space equivalents can be viewed
as a deformation of the point product in the algebra $\mathcal A $ of $C^\infty$ phase-space functions with the  Gr\"oenewold-Moyal multi-differential operator:
\be \star_{\hbar}:=\exp [ \frac{i\hbar}{2} \Lambda ] :=
\exp \left[ \frac{i\hbar}{2} ( {\overleftarrow{\nabla_{{\bf
q}}}}\cdot {\overrightarrow{\nabla_{{\bf p}}}} -
{\overleftarrow{\nabla_{{\bf p}}}}\cdot
{\overrightarrow{\nabla_{{\bf q}}}} ) \right]\label{2.11} \ee
inducing this deformation. This concept of a twisted product was given a more general mathematical framework by Bayen {\it et al}
in \cite{bayen}, whose proposed  deformation quantization paradigm and noncommutative symbol calculus, led to an autonomous reformulation
of quantum theory directly in terms of phase-space functions,
composed via the twisted or $\star$-product, instead of operators and Hilbert space states.

While applications of the original Weyl-Wigner-Gr\"oenewold-Moyal (WWGM) formalism were restricted to the description of systems in flat phase space,
the systems under consideration in the more general deformation quantization scheme possess an intrinsic group of symmetries, with the phase-space being an
homogeneous manifold on which the group of transformations acts transitively \cite{frons, huynh, basart, moreno, antonsen}. This implies
then the possibility of extending the phase space approach to the ``quantization" of curved spaces. However,
for the various known versions of deformation theory there are a large variety of $\star$-products which in turn imply, in general, different quantum mechanical theories for the same problem.\\
In order to deal with such non-uniqueness and arrive at a $\star$-product that would ensure the physical equivalence of deformation quantization with the ordinary
quantum mechanics, the need for supplementary conditions has been suggested, so that the linear bijective mapping between operators on Hilbert
space and classical functions on phase space can be implemented by a kernel operator which satisfies a number of physically sensible postulates thus hopefully providing a scheme to single out the most
adequate symbol calculus from the many that have and could be proposed. \\
Moreover, such non-uniqueness becomes manifest even for quantum deformation schemes with known equivalent $\star$-products in flat space-time standard quantum mechanics, when space-space and/or space-time non-commutativity is incorporated into the formalism. This noncommutative quantum mechanics and the behavior of classical fields, defined as functions of noncommutative spatial variables, has been
the object of a great deal of attention in the last years. Physicists became attracted to the more mathematical aspects of deformation quantization with the hope that such theories would provide the tools needed to remove the singularities in physical field theories without the need of renormalization. Although these expectations have not materialized up to now, noncommutative field theory and its quantum mechanical mini-superspace have led to many new and interesting results. In particular, in the context of string theory there has been a lot of interest in studying solitonic solutions of noncommutative field theory \cite{gopa, volo}. Also motivated by that work, but in a somewhat different direction,
 coherent structures in the form of noncommutative solitons and vortices were studied by the authors in a recent collaboration \cite{tim}. It was shown there that the noncommutativity of the spatial variables, when averaged with vortex or plateau type coherent states, induced an effective lattice structure of Landau cells whose distribution and size depended on the coherent states considered. This shows that the effect of the noncommutativity on coherent structures, with an amplitude comparable to the scale parameter  $\theta$ of noncommutativity of the $\star$-product,
is to induce a behavior of classical structures in a physical lattice whose dynamics can be described in terms of a Peierls-Nabarro potential. It would not be unreasonable to expect that such dynamical creation of lattice structures as an effect of the noncommutativity on coherent states, which mathematically would be reflected in the replacement of differential field equations by equations of differences, could be related to another important quantization scheme known as loop quantum gravity. This final objective forms part of an ongoing program initiated in \cite{tim}, and it is within that much wider context that the present work is intended. \\
Thus, in order to arrive at an identification of the $\star$-product appropriate for the above mentioned program, we will here specifically start by extending the WWGM procedure in order to analyze a space-space noncommutative Heisenberg-Weyl algebra (again, noncommutativity being understood here as a non-vanishing commutator between the operators of spatial coordinates or momenta) in order to obtain the generalization of the well known expressions of the Heisenberg-Weyl algebra of usual Quantum Mechanics. Afterwards we will apply to this same Lie algebra two quantization formalisms which are purportedly more general and that were developed to provide a quantization scheme even for curved spaces. The first one started with the work of Stratonovich \cite{strat} and was further developed elsewhere \cite{vari, brif}. The second corresponds to the Berezin geometric quantization program of covariant and contravariant symbols for K\"ahler manifolds \cite{bere}. Finally we derive the additional specific requirements that need to be imposed on these different schemes, in order to obtain $\star$-valued equations which constitute a stronger quantization requirement, as they relate eigenvalues of the physical states appearing in the density matrix to the Weyl equivalents of the operator observables.\\

\section{The WWGM Phase-space Quantum Mechanics Based on the space-space noncommutative Heisenberg-Weyl Lie Algebra }
By a space-space (and/or momentum-momentum) noncommutative Heisenberg-Weyl algebra we understand \cite{rosen} the algebra of
position and momentum operators satisfying the commutation relations
\begin{align}
[{\hat R}_i, {\hat R}_j]  &= i\theta_{ij}\hat I \nonumber\\
[{\hat P}_i, {\hat P}_j] &= i\hbar {\bar \theta}_{ij}\hat I\label{noncomm2} \\
[{\hat R}_i,{\hat P}_j] &= i \hbar \delta_{ij}\hat I \nonumber
\end{align}
where ${\hat R}_i$,${\hat P}_i$\;\; $i=1,\dots,d$  are the
components of the position and momentum quantum operators,
respectively, with component eigenvalues on ${\mathbb R}^{d}$, 
the identity $\hat I$ is the central element of the algebra, and
$\theta_{ij}$ and ${\bar \theta}_{ij}$ are evidently antisymmetric
matrices, which in the most general case can be functions of the
generators of the above algebra. For our present purposes and
algebraic simplicity, in what follows we shall set ${\bar
\theta}_{ij}=0$ and $d=2$, and consider only the zeroth order
constant term of the Taylor expansion of $\theta_{12}\equiv \theta$.

From an intrinsically noncommutative operator point of view, the
development of a formulation for the quantum mechanics based on the
above Heisenberg-Weyl algebra of operators requires first a
specification of a representation for the generators of the algebra,
second a specification of the Hamiltonian which governs the time
evolution of the system and last a specification of the Hilbert
space on which these operators and the other observables of the
theory act. As for the choice of the Hilbert space, a reasonable
assumption is that it can be taken to be the same as that for the
corresponding system in the usual quantum mechanics, but for a
realization of the space-space noncommutative Heisenberg-Weyl algebra, because of the
noncommutativity (\ref{noncomm2}), we can not use configuration
space as a basis. We can use, however, for a basis either of the
eigenkets $|p_1,p_2\rangle$, $|q_1,p_2 \rangle$, $|q_2,p_1 \rangle$,
of the commuting pairs of observables $({\hat P}_1,{\hat P}_2)$,
$({\hat R}_1,{\hat P}_2)$, or $({\hat R}_2,{\hat P}_1)$,
respectively, or any combination of the
$(R,P)$ such that they form a complete set of commuting observables. \\
 Specifically, we choose as the realization of our
Heisenberg-Weyl algebra the one based on $|q_1,p_2 \rangle$. The
construction follows standard procedures ({\it cf. e.g.} \cite{messiah}) and
it is detailed in \cite{rosen}.
We then have that ${\hat R}_2$ in this basis is realized by
\be {\hat R}_2 =-i\theta\partial_{q_1}
+ i\hbar\partial_{p_2}\label{2.2} \ee
and
\be {\hat P}_1 = -i\hbar\partial_{q_1}\label{2.3}
\ee The representations for the remainder of the generators ${\hat
R}_1$ and ${\hat P}_2$ of the algebra are obviously just
multiplicative.
Note that the change of basis
$|q_1,p_2\rangle\rightarrow|q_2,p_1\rangle$ follows directly from the
transition function $\langle q_1,p_2|q_2,p_1\rangle$, which is
derived \cite{acat} by noting that \be \langle q_1,p_2|{\hat R}_2
|q_2,p_1\rangle = q_2 \langle q_1,p_2|q_2,p_1\rangle
=i(\hbar\partial_{p_2} -\theta\partial_{q_1})\langle
q_1,p_2|q_2,p_1\rangle\label{2.4} \ee and \be \langle q_1,p_2|{\hat
P}_1 |q_2,p_1\rangle = p_1 \langle q_1,p_2|q_2,p_1\rangle
=-i\hbar\partial_{q_1}\langle q_1,p_2|q_2,p_1\rangle\label{2.5} \ee
Combining these two expressions yields \be (\hbar q_2 - \theta
p_1)\langle q_1,p_2|q_2,p_1\rangle=i\hbar\partial_{p_2} \langle
q_1,p_2|q_2,p_1\rangle\label{2.6} \ee which can be readily solved
to give, after normalization,
\be
\langle q_1,p_2|q_2,p_1\rangle=\frac{1}{2\pi\hbar}\exp[-\frac{i}{\hbar}(q_2
p_2 - \frac{\theta}{\hbar} p_1 p_2 -q_1 p_1)]\label{2.7} \ee

Since the displacement operators $\{(2\pi\hbar)^{-1} \exp[\frac{i}{\hbar}({\bf y}\cdot {\bf\hat  R}
+{\bf x}\cdot{\bf\hat  P})]\}$, where ${\bf x}= (x_1, x_2)\;\;\; {\bf y}= (y_1, y_2)$, form a complete orthonormal set in the space-space noncommutative Heisenberg algebra
any Schr\"{o}dinger operator (which may depend
explicitly on time) $A({\bf \hat P}, {\bf \hat R}, t)$ can be
written as \be A({\bf \hat P}, {\bf \hat R}, t)= \int \int d{\bf x}
\ d{\bf y} \alpha({\bf x}, {\bf y}, t) \exp[\frac{i}{\hbar}({\bf
x}\cdot {\bf \hat P} + {\bf y}\cdot {\bf \hat R})]\label{op} \ee
where
the $c$-function $\alpha({\bf x}, {\bf y},
t)$ is determined by \be \alpha({\bf x}, {\bf y},
t)=(2\pi\hbar)^{-2} {\rm Tr} \{ A({\bf \hat P}, {\bf \hat R}, t)
\exp[-\frac{i}{\hbar}({\bf x}\cdot {\bf \hat P}
 + {\bf y}\cdot {\bf \hat R})] \}\label{alpha}
\ee The Weyl function corresponding to the quantum operator $A({\bf
\hat P}, {\bf \hat R}, t)$ is then given by 
\be\begin{split}
W_{A} ({\bf p},{\bf q}, t)= \int \int d{\bf x} \ d{\bf y}
\;\alpha({\bf x},
{\bf y}, t) \exp[\frac{i}{\hbar}({\bf x}\cdot {\bf p} + {\bf y}\cdot {\bf q})]=\hspace{1in}\\
\int\int dx_1 dy_2 e^{\frac{i}{\hbar}(x_1 p_1 +y_2 q_2)}\langle q_1
-\frac{x_1}{2}-\frac{\theta y_2}{2\hbar}, p_2 +\frac{y_2}{2}|{\hat
A}|q_1 +\frac{x_1}{2}+\frac{\theta y_2}{2\hbar}, p_2
-\frac{y_2}{2}\rangle \label{weyl}
\end{split}\ee
To derive the expectation value of a product of two
Schr\"{o}dinger operators, one writes the expectation value of the
product in terms of the von Neumann density matrix ${\boldsymbol
\rho}$  as \be \langle{\hat A}_1 {\hat A}_2 \rangle={\rm
Tr}[{\boldsymbol \rho}{\hat A}_1 {\hat A}_2]\label{2.9} \ee and
evaluates the trace in the above chosen basis. Thus by using
completeness of the basis $|q_1, p_2\rangle$ and substituting
(\ref{op}) for the operators ${\hat A}_1$ and ${\hat A}_2$, equation (\ref{2.9}) then becomes
\be
\begin{split}\langle{\hat A}_1 {\hat A}_2
\rangle=\int d{\bf x}d{\bf y}d{\bf u}d{\bf
v}dq_1dp_2dq'_1dp'_2dq''_1dp''_2\langle q_1,p_2|{\boldsymbol
\rho}|q'_1,p'_2\rangle\alpha_1({\bf x},{\bf y},t)\alpha_2({\bf
u},{\bf v},t)\\ 
(\times)\langle
q'_1,p'_2|e^{\frac{i}{\hbar}({\bf x}\cdot {\bf \hat P} + {\bf
y}\cdot {\bf \hat R})}|q''_1,p''_2\rangle\langle
q''_1,p''_2|e^{\frac{i}{\hbar}({\bf u}\cdot {\bf \hat P} + {\bf v}\cdot {\bf \hat R})}|q_1,p_2\rangle
\end{split}
\ee

Moreover, resorting to the Baker-Campbell-Hausdorff theorem, making use of (\ref{2.7}) and performing the integrals over
$q'_1,p'_2,q''_1$ and $p''_2$ we obtain\\

\begin{align}\label{2.16}\langle{\hat A}_1 {\hat A}_2 \rangle=&\int d{\bf
x}d{\bf y}d{\bf u}d{\bf v}dq_1dp_2\langle q_1,p_2|{\boldsymbol
\rho}|q_1-x_1-u_1-\frac{v_2\theta}{\hbar}-\frac{y_2\theta}{\hbar},p_2+y_2+v_2\rangle\alpha_1({\bf
x},{\bf y},t)\alpha_2({\bf u},{\bf v},t)\nonumber\\
&(\times)\exp[\frac{i}{\hbar}(y_1q_1-y_1u_1+v_1q_1+x_2p_2+x_2v_2+u_2p_2-\frac{y_1x_1}{2}+\frac{y_2x_2}{2}
-\frac{v_1u_1}{2}+\frac{u_2v_2}{2})]\nonumber\\
&(\times)\exp[\frac{i}{\hbar}(-\frac{\theta}{\hbar}y_1v_2-\frac{\theta}{2\hbar}y_1y_2
-\frac{\theta}{2\hbar}v_1v_2)].\end{align}
Making now the change of
variables $q_1=\xi,p_2=\eta$ and substituting $\alpha_1({\bf x},{\bf
y},t)$ and $\alpha_2({\bf u},{\bf v},t)$ in terms of their
corresponding Weyl functions, equation (\ref{2.16}) becomes
\begin{align}\langle{\hat A}_1 {\hat A}_2
\rangle=\left(\frac{1}{2\pi\hbar}\right)^{8}&\int d{\bf p}d{\bf
q}d{\bf p'}d{\bf q'}d{\bf x}d{\bf y}d{\bf u}d{\bf v}d\xi
d\eta\langle \xi,\eta|{\boldsymbol
\rho}|\xi-x_1-u_1-\frac{v_2\theta}{\hbar}-\frac{y_2\theta}{\hbar},\eta+y_2+v_2\rangle\nonumber\\
&(\times)W_{A_1}({\bf p},{\bf q},t)W_{A_2}({\bf p'},{\bf
q'},t)\exp[\frac{i}{\hbar}y_1(\xi-u_1-\frac{\theta}{\hbar}v_2-\frac{x_1}{2}-\frac{\theta}{2\hbar}y_2-q_1)]\nonumber\\
&(\times)\exp[\frac{i}{\hbar}v_1(\xi-\frac{u_1}{2}-\frac{\theta}{2\hbar}v_2-q_1')]e^{\frac{i}{\hbar}v_2(x_2+\frac{u_2}{2}-q'_2)}e^{\frac{i}{\hbar}y_2(\frac{x_2}{2}-q_2)}\nonumber\\
&(\times)e^{-\frac{i}{\hbar}x_1p_1}e^{-\frac{i}{\hbar}u_1p'_1}e^{-\frac{i}{\hbar}x_2(p_2-\eta)}e^{-\frac{i}{\hbar}u_2(p'_2-\eta)}.
\label{2.17}\end{align}

Next we integrate over
$y_1,x_2,v_1,u_2,u_1,v_2,\xi$ and $\eta$ to get
\begin{align}\langle{\hat A}_1 {\hat A}_2
\rangle=\frac{4}{(2\pi\hbar)^{4}}&\int d{\bf p}d{\bf q}d{\bf
p'}d{\bf q'}dx_1dy_2\langle 2q'_1-q_1-\frac{x_1}{2}-\frac{\theta
y_2}{2\hbar},2p'_2-p_2+\frac{y_2}{2}|{\boldsymbol
\rho}|q_1-\frac{x_1}{2}-\frac{\theta
y_2}{2\hbar},p_2+\frac{y_2}{2}\rangle\nonumber\\
&(\times)W_{A_1}({\bf p},{\bf q},t)W_{A_2}({\bf p'},{\bf
q'},t)e^{-\frac{i}{\hbar}y_2q_2}e^{-\frac{i}{\hbar}x_1p_1}\nonumber\\
&(\times)e^{-\frac{i}{\hbar}q'_2(2p_2-2p'_2-y_2)}e^{-\frac{i}{\hbar}p'_1(2q'_1-2q_1-\frac{2\theta}{\hbar}p_2+\frac{2\theta}{\hbar}p'_2-x_1)}
\label{2.18}\end{align}

Observe now that this expression can also be written as
\begin{align}\langle{\hat A}_1 {\hat A}_2
\rangle=\frac{4}{(2\pi\hbar)^{4}}&\int d{\bf p}d{\bf q}d{\bf
p'}d{\bf q'}dx_1dy_2\left[e^{\frac{\theta
y_2}{\hbar}\partial_{x_1}}\langle
2q'_1-q_1-\frac{x_1}{2},2p'_2-p_2+\frac{y_2}{2}|{\boldsymbol
\rho}|q_1-\frac{x_1}{2},p_2+\frac{y_2}{2}\rangle\right]\nonumber\\
&(\times)W_{A_1}({\bf p},{\bf q},t)W_{A_2}({\bf p'},{\bf
q'},t)e^{-\frac{i}{\hbar}y_2q_2}e^{-\frac{i}{\hbar}x_1p_1}\nonumber\\
&(\times)e^{-\frac{i}{\hbar}q'_2(2p_2-2p'_2-y_2)}e^{-\frac{i}{\hbar}p'_1(2q'_1-2q_1-\frac{2\theta}{\hbar}p_2+\frac{2\theta}{\hbar}p'_2-x_1)}
\end{align}and after integrating by parts we obtain
\begin{align}\label{intparts}\langle{\hat A}_1 {\hat A}_2
\rangle=\frac{4}{(2\pi\hbar)^{4}}&\int d{\bf p}d{\bf q}d{\bf
p'}d{\bf q'}dx_1dy_2\langle
2q'_1-q_1-\frac{x_1}{2},2p'_2-p_2+\frac{y_2}{2}|{\boldsymbol
\rho}|q_1-\frac{x_1}{2},p_2+\frac{y_2}{2}\rangle\nonumber\\
&(\times)W_{A_1}({\bf p},{\bf q},t)W_{A_2}({\bf p'},{\bf
q'},t)e^{-\frac{i}{\hbar}y_2q_2}e^{-\frac{i}{\hbar}q'_2(2p_2-2p'_2-y_2)}e^{-\frac{i}{\hbar}p'_1(2q'_1-2q_1)}\nonumber\\
&(\times)e^{\frac{i}{\hbar}x_1(p'_1-p_1)}e^{-\frac{i}{\hbar^{2}}\theta
y_2(p'_1-p_1)}e^{\frac{2i}{\hbar^{2}}\theta p'_1(p_2-p'_2)}
\end{align}
To reconstruct the star product that should
arise from this formulation we use the following identities
\begin{align}e^{-\frac{\theta}{\hbar}p'_1\partial_{q'_2}}e^{\frac{i}{\hbar}q'_2y_2}&=
e^{\frac{i}{\hbar}q'_2y_2}e^{-\frac{i\theta}{\hbar^2}y_2p'_1}\nonumber\\
e^{-\frac{\theta}{\hbar}p_1\partial_{q_2}}e^{-\frac{i}{\hbar}q_2y_2}&=
e^{-\frac{i}{\hbar}q_2y_2}e^{\frac{i\theta}{\hbar^2}y_2p_1}\\
e^{-\frac{\theta}{\hbar}p'_1\partial_{q'_2}}e^{-\frac{2i}{\hbar}q'_2(p_2-p'_2)}&=
e^{-\frac{2i}{\hbar}(p_2-p'_2)(q'_2-\frac{\theta}{\hbar}p'_1)}\nonumber\end{align}
so that (\ref{intparts}) becomes
\begin{align}\langle{\hat A}_1 {\hat A}_2
\rangle=\frac{4}{(2\pi\hbar)^{4}}&\int d{\bf p}d{\bf q}d{\bf
p'}d{\bf q'}dx_1dy_2\langle
2q'_1-q_1-\frac{x_1}{2},2p'_2-p_2+\frac{y_2}{2}|{\boldsymbol
\rho}|q_1-\frac{x_1}{2},p_2+\frac{y_2}{2}\rangle\nonumber\\
&(\times)W_{A_1}({\bf p},{\bf q},t)W_{A_2}({\bf p'},{\bf
q'},t)e^{-\frac{i}{\hbar}p'_1(2q'_1-2q_1)}e^{\frac{i}{\hbar}x_1(p'_1-p_1)}\nonumber\\
&(\times)e^{-\frac{\theta}{\hbar}p'_1\partial_{q'_2}}\left(e^{\frac{i}{\hbar}q'_2y_2}e^{-\frac{2i}{\hbar}q'_2(p_2-p'_2)}\right)
\left(e^{-\frac{\theta}{\hbar}p_1\partial_{q_2}}e^{-\frac{i}{\hbar}q_2y_2}\right)
\end{align}
After integrating by parts the above equation reads
\begin{align}\label{2.22}\langle{\hat A}_1 {\hat A}_2
\rangle=\frac{4}{(2\pi\hbar)^{4}}&\int d{\bf p}d{\bf q}d{\bf
p'}d{\bf q'}dx_1dy_2\langle
2q'_1-q_1-\frac{x_1}{2},2p'_2-p_2+\frac{y_2}{2}|{\boldsymbol
\rho}|q_1-\frac{x_1}{2},p_2+\frac{y_2}{2}\rangle\nonumber\\
&(\times)W_{A_1}({\bf
p},q_1,q_2+\frac{\theta}{\hbar}p_1,t)W_{A_2}({\bf
p'},q'_1,q'_2+\frac{\theta}{\hbar}p'_1,t)e^{-\frac{i}{\hbar}p'_1(2q'_1-2q_1)}e^{\frac{i}{\hbar}x_1(p'_1-p_1)}\nonumber\\
&(\times)e^{\frac{i}{\hbar}y_2(q'_2-q_2)}e^{-\frac{2i}{\hbar}q'_2(p_2-p'_2)}
\end{align}
Now make the following change of variables
\begin{align}\label{2.23}x_1=2q_1-2z_1,&\quad y_2=2z_2-2p_2\nonumber\\
q'_1=q_1+\mu_1,&\quad q'_2=q_2+\mu_2 \nonumber\\
p'_1=p_1+\nu_1,&\quad p'_2=p_2+\nu_2\end{align}
to obtain
\begin{align}\label{2.24}\langle{\hat A}_1 {\hat A}_2
\rangle=\frac{16}{(2\pi\hbar)^{4}}&\int d{\bf p}d{\bf
q}d\mu_1d\mu_2d\nu_1d\nu_2dz_1dz_2\langle
z_1+2\mu_1,z_2+2\nu_2|{\boldsymbol \rho}|z_1,z_2\rangle
e^{-\frac{2i}{\hbar}\mu_1p_1}e^{\frac{2i}{\hbar}\nu_2q_2}\nonumber\\
&(\times)e^{-\frac{2i}{\hbar}\nu_1(\mu_1-q_1+z_1)}e^{-\frac{2i}{\hbar}\mu_2(p_2-\nu_2-z_2)}W_{A_1}({\bf
p},q_1,q_2+\frac{\theta}{\hbar}p_1,t)\nonumber\\
&(\times)e^{\nu_1\partial_{p_1}}e^{\nu_2\partial_{p_2}}e^{\mu_1\partial_{q_1}}e^{\mu_2\partial_{q_2}}W_{A_2}
({\bf p},q_1,q_2+\frac{\theta}{\hbar}p_1,t)\nonumber\\.
\end{align}
But
\begin{align}
e^{\frac{2i}{\hbar}q_1\nu_1}e^{\nu_1\overrightarrow{\partial}_{p_1}}W_{A_2}&=e^{\frac{2i}{\hbar}q_1\nu_1}e^{-\frac{i\hbar}{2}
\overleftarrow{\partial}_{q_1}\overrightarrow{\partial}_{p_1}}W_{A_2}\nonumber\\
e^{\frac{2i}{\hbar}q_2\nu_2}e^{\nu_2\overrightarrow{\partial}_{p_2}}W_{A_2}&=e^{\frac{2i}{\hbar}q_2\nu_2}e^{-\frac{i\hbar}{2}
\overleftarrow{\partial}_{q_2}\overrightarrow{\partial}_{p_2}}W_{A_2}\label{2.28}\\
e^{-\frac{2i}{\hbar}p_1\mu_1}e^{\mu_1\overrightarrow{\partial}_{q_1}}W_{A_2}&=e^{-\frac{2i}{\hbar}p_1\mu_1}e^{\frac{i\hbar}{2}
\overleftarrow{\partial}_{p_1}\overrightarrow{\partial}_{q_1}}W_{A_2}\nonumber\\
e^{-\frac{2i}{\hbar}p_2\mu_2}e^{\mu_2\overrightarrow{\partial}_{q_2}}W_{A_2}&=e^{-\frac{2i}{\hbar}p_2\mu_2}e^{\frac{i\hbar}{2}
\overleftarrow{\partial}_{p_2}\overrightarrow{\partial}_{q_2}}W_{A_2}\nonumber\end{align}
which, when substituted into (\ref{2.24}) and
integrated by parts results in
\begin{align}\label{2.29}\langle{\hat A}_1 {\hat A}_2
\rangle=\frac{16}{(2\pi\hbar)^{4}}&\int d{\bf p}d{\bf
q}d\mu_1d\mu_2d\nu_1d\nu_2dz_1dz_2\langle
z_1+2\mu_1,z_2+2\nu_2|{\boldsymbol \rho}|z_1,z_2\rangle
e^{-\frac{2i}{\hbar}\mu_1p_1}e^{\frac{2i}{\hbar}\nu_2q_2}\nonumber\\
&(\times)e^{-\frac{2i}{\hbar}\nu_1(\mu_1-q_1+z_1)}e^{-\frac{2i}{\hbar}\mu_2(p_2-\nu_2-z_2)}\nonumber\\
&(\times)\left[W_{A_1}({\bf
p},q_1,q_2+\frac{\theta}{\hbar}p_1,t)\star_{\hbar}W_{A_2}({\bf
p},q_1,q_2+\frac{\theta}{\hbar}p_1,t)\right]\end{align}
Last, integrating over $\nu_1,\mu_2,\mu_1$ and $\nu_2$ and performing the final
change of variables \\
$z_1=q_1+\frac{s_1}{2},z_2=p_2+\frac{s_2}{2}$,
equation (\ref{2.29}) takes the form\\
\begin{align}\label{2.30}\langle{\hat A}_1 {\hat A}_2
\rangle=\frac{1}{(2\pi\hbar)^{2}}&\int d{\bf p}d{\bf
q}ds_1ds_2\langle q_1-\frac{s_1}{2},p_2-\frac{s_2}{2}|{\boldsymbol
\rho}|q_1+\frac{s_1}{2},p_2+\frac{s_2}{2}\rangle
e^{\frac{i}{\hbar}s_1p_1}\nonumber\\
&(\times)e^{-\frac{i}{\hbar}s_2q_2}\left[W_{A_1}({\bf
p},q_1,q_2+\frac{\theta}{\hbar}p_1,t)\star_{\hbar}W_{A_2}({\bf
p},q_1,q_2+\frac{\theta}{\hbar}p_1,t)\right]\end{align}

Recalling the definition of the Wigner function:
\be\label{2.31}\rho_w({\bf p},{\bf q}):=\frac{1}{(2\pi\hbar)^{2}}\int
ds_1ds_2\langle q_1-\frac{s_1}{2},p_2-\frac{s_2}{2}|{\boldsymbol
\rho}|q_1+\frac{s_1}{2},p_2+\frac{s_2}{2}\rangle
e^{\frac{i}{\hbar}s_1p_1}e^{-\frac{i}{\hbar}s_2q_2}\ee
equation (\ref{2.30}) may be expressed in the compact form
\be\label{starproduct}\langle{\hat A}_1 {\hat A}_2
\rangle=\int d{\bf p}d{\bf q}
\rho_w({\bf p},{\bf q})\left[W_{A_1}({\bf
p},q_1,q_2+\frac{\theta}{\hbar}p_1,t)\star_{\hbar}W_{A_2}({\bf
p},q_1,q_2+\frac{\theta}{\hbar}p_1,t)\right]\ee
where
\be\label{qstar}
\star_\hbar := \exp\left [\sum_{i=1,2} \frac{i\hbar}{2}(\overleftarrow \partial_{q_i}\overrightarrow\partial_{p_i}
-\overleftarrow\partial_{p_i}\overrightarrow\partial_{q_i})\right]
\ee
Consequently, in the
phase-space formulation of Quantum Mechanics based on the 
algebra (\ref{noncomm2}), the algebra of Weyl functions is deformed by a
$\star$-product defined by
\be\label{starproduct2}
W_{A_1}\star W_{A_2}:=
m\circ [e^{\sum_{i=1,2}\frac{i\hbar}{2}(\partial_{q_i}\otimes\partial_{p'_i}-\partial_{q'_i}\otimes\partial_{p_i})}
\circ
e^{\frac{\theta}{\hbar}p_1\partial_{q_2}}\otimes e^{\frac{\theta}{\hbar}p'_1
\partial_{q'_2}}W_{A_1}({\bf p},{\bf q})\otimes W_{A_2}({\bf p'},{\bf q'})]_{\bf{q,p=q',p'}}
\ee

In addition, by a similar calculation to the one above, we can show that the Weyl symbol
\be\label{weylwig}
W_{\rho}({\bf p},{\bf q})=(2\pi\hbar)^{-2} \int d{\bf x}d{\bf y} {\rm Tr}[\boldsymbol\rho e^{-\frac{i}{\hbar}({\bf x}\cdot {\bf P}+ {\bf y}\cdot {\bf R})}]
e^{\frac{i}{\hbar}({\bf x}\cdot {\bf p}+ {\bf y}\cdot {\bf q})}
\ee
associated with the density matrix $\boldsymbol\rho$ is related to the Wigner function by

\be\label{weylwig3}
W_{\rho}({\bf p},{\bf q})=e^{-\frac{\theta}{\hbar}p_1 \partial_{q_2}} \rho_w({\bf p},{\bf q})
\ee
Hence for the space-space noncommutative Heisenberg-Weyl algebra, the Weyl symbol of the density matrix and the Wigner function
as defined in (\ref{2.31}) are not the same, contrary from what is the case for the usual quantum mechanics Heisenberg algebra;
{\it i.e.}
\be\label{lim}
 W_{\rho}({\bf p},{\bf q}) \stackrel{\theta\to 0}{\longrightarrow} \rho_w({\bf p},{\bf q}) .
\ee\\

Note now that if we substitute (\ref{weylwig3}) into (\ref{starproduct}) and integrate by parts, we get\\

\begin{eqnarray}\label{starproduct3}
\langle{\hat A}_1 {\hat A}_2\rangle&=&\int d{\bf p}d{\bf q}
W_{\rho}({\bf p},{\bf q})e^{-\frac{\theta}{\hbar}p_1 \overrightarrow{\partial}_{q_2}}\left[W_{A_1}({\bf
p},q_1,q_2+\frac{\theta}{\hbar}p_1,t)\star_{\hbar}W_{A_2}({\bf
p},q_1,q_2+\frac{\theta}{\hbar}p_1,t)\right]\nonumber\\
&=& \int d{\bf p}d{\bf q}W_{\rho}({\bf p},{\bf q})
e^{-\frac{\theta}{\hbar}p_1 \overrightarrow{\partial}_{q_2}}\nonumber\\
&\times&\left[W_{A_1}\left(p_1 - \frac{i\hbar}{2}\overrightarrow{\partial}_{q_1},
p_2 - \frac{i\hbar}{2}\overrightarrow{\partial}_{q_2}, q_1,q_2+\frac{i\hbar}{2}\overrightarrow{\partial}_{p_2}
+\frac{\theta}{\hbar}(p_1 -\frac{i\hbar}{2}\overrightarrow{\partial}_{q_1}),t\right)\right.\\
&\times& \left. W_{A_2}\left({\bf p},q_1 -\frac{i\hbar}{2}\overleftarrow{\partial}_{p_1},q_2 +\frac{\theta}{\hbar}p_1,t\right)\right]\nonumber\\
&=&\int d{\bf p}d{\bf q} W_{\rho}({\bf p},{\bf q})
\left[W_{A_1}({\bf p},{\bf q},t)\star_{\theta}\circ\star_\hbar W_{A_2}({\bf
p},{\bf q},t)\right]\nonumber
\end{eqnarray}
where
\be\label{starproduct4}
\star_\theta \circ \star_\hbar:= e^{\frac{i\theta}{2}(\overleftarrow{\partial}_{q_1} \overrightarrow{\partial}_{q_2}
-\overleftarrow{\partial}_{q_2} \overrightarrow{\partial}_{q_1})} \circ
\exp\left [\sum_{i=1,2} \frac{i\hbar}{2}(\overleftarrow \partial_{q_i}\overrightarrow\partial_{p_i}-\overleftarrow\partial_{p_i}\overrightarrow\partial_{q_i})\right]
\ee
Clearly the expectation values obtained from (\ref{2.30}) and (\ref{starproduct3}) are the same. However, since for the space-space noncommutative Heisenberg-Weyl algebra
the Wigner function associated with the density matrix $\hat{\boldsymbol\rho}$ and its corresponding Weyl symbol are not the same, the
twistings in (\ref{weylwig}) and (\ref{starproduct3}) of the product of Weyl symbols of two arbitrary operators do not agree in general. Their explicit forms are obviously
basis dependent as well as
dependent on whether averaging is done relative to the Wigner function or the Weyl symbol of the density matrix.

Furthermore, given the two different $\star$-products (\ref{starproduct2}) and  (\ref{starproduct4}) of a pair of Weyl-symbols,
it is pertinent to inquire which of them corresponds to the Weyl-symbol of a product of two operators. To answer this
question univocally we need to make use of (\ref{2.7}), (\ref{alpha}) and (\ref{weyl}). After a rather lengthy but fairly direct calculation
one can show that
\be\label{weylprod}
W_{A_1 A_2} = W_{A_1}\star_\theta \circ \star_\hbar W_{A_2}
\ee
So, for the quantum mechanics based on the space-space noncommutative Heisenberg-Weyl Lie group, we need to make iterative use of (\ref{weylprod})
for the calculation of Weyl-symbols corresponding to quantum operators. In particular, note that the Weyl-symbol
corresponding to an operator $\hat A_1 =\hat A_1(\bf \hat P)$ which is a function only of the momenta operators is
given by the c-function $W_{A_1}(\bf p)$ having the same functional form as the quantum operator, as it is the case in the usual
WWGM quantum mechanics. On the other hand, for q-functions of the position operators this is not always true for the space-space noncommutative Heisenberg-Weyl group,
as can be easily seen, when consider for example the Weyl-symbol associated with the operator $\hat R_1 \hat R_2$, for which (\ref{weylprod})
yields $W_{R_1 R_2}= (q_1 +i\frac{\theta}{2}\partial_{q_2})q_2 =q_1 q_2 + i\frac{\theta}{2}$.

From a statistical point of view, both the Wigner function
(\ref{2.31}) and the Weyl symbol (\ref{weylwig}) for the density matrix admit a quasi-probabilistic interpretation,
although the projected density probabilities  are not all the same. Indeed, projecting (\ref{2.31}) onto the plane
$q_1 - p_2$ ({\it i.e.} integrating over $q_2, p_1$) immediately yields
\be\label{proj1}
\int dp_1 dq_2 \rho_w ({\bf p, \bf q}) = \langle q_1, p_2| \hat{\boldsymbol \rho}|q_1, p_2\rangle
\ee
while projecting onto the $q_2 - p_1$ plane by making use of (\ref{2.7}) results in
\be\label{proj2}
\int dp_2 dq_1 \rho_w ({\bf p, \bf q}) = \langle q_2 +(\theta/\hbar) p_1, p_1| \hat{\boldsymbol \rho}|q_2 +(\theta/\hbar) p_1, p_1\rangle
\ee
On the other hand, if we perform the same calculations for the corresponding
Weyl symbol, we find
\be\label{proj3}
\int dp_1 dq_2 W_{\rho} ({\bf p, \bf q}) = \langle q_1 , p_2| \hat{\boldsymbol \rho}|q_1 , p_2\rangle
\ee
\be\label{proj4}
\int dp_2 dq_1 W_{\rho} ({\bf p, \bf q}) = \langle q_2 , p_1| \hat{\boldsymbol \rho}|q_2, p_1\rangle
\ee

Let us now see how the above results compare with the ones resulting from applying the
Stratonovich-Weyl Correspondence and the Berezin geometric quantization to the space-space noncommutative Heisenberg-Weyl Lie group.

\section{The Stratonovich-Weyl Correspondence for the space-space noncommutative Heisenberg-Weyl Lie Group }

In order to make our discussion self-contained and fix notation, we begin by summarizing
the essential elements of the Stratonovich-Weyl correspondence. For a considerably more ample presentation of this formalism
we refer the reader to the work in references \cite{strat}, \cite{vari} and \cite{brif}.

Let $X$ be an even dimensional homogeneous space given by the quotient $G/H$, where $G$ is a simply connected Lie group
(of finite dimension $n$) describing the dynamical symmetry  of a given quantum system, and $H\subset G$
its isotropy subgroup . If $X$ is given a K\"ahlerian structure, then it can be interpreted as the phase space of
a classical dynamical system. The mapping $\Omega \to |\Omega\rangle\langle\Omega|, $ where $\Omega=\Omega(g)$ is a point in $X$ and
 $g\in G$, is the geometric quantization for this system \cite{bere}. \\
The Stratonovich generalization of the standard Gr\"oenewold-Moyal quantization to quantum systems possessing and intrinsic
group $G$ of symmetries is based on the following postulates:\\
(i) Linearity: There is a one-to-one map $\hat A \rightarrow W_{A}(\Omega)$, .\\
(ii) Reality: $W_{A^{\dag}}(\Omega)=[W_{A}(\Omega)]^\ast$.\\
(iii) Standardization: $\int_X d\mu(\Omega) \: W_{A}(\Omega)={\rm Tr} \:\hat A$, where $d\mu(\Omega)$ is the invariant space measure.\\
(iv) Traciality: $\int_X d\mu(\Omega)  W_{A_1}(\Omega)\: W_{A_2}(\Omega)={\rm Tr} (\:\hat A_1 \hat A_2)$.\\
(v) Covariance: $W_{g\cdot A}(\Omega)=W_{A}(g^{-1}\cdot\Omega)$, where $g\cdot A$ denotes the adjoint action of a unitary irreducible\\
   representation $\pi$ of $G$ on $\hat A$.

A function $W_{A}(\Omega)$ satisfying these five properties is known as the Stratonovich-Weyl (SW) symbol
associated with a quantum operator $\hat A$ acting on Hilbert space. The linearity map is implemented
by means of the generalized Weyl rule
\be
W_{A}(\Omega)={\rm Tr} \:[\hat A \: \Delta(\Omega)] \label{lin}
\ee
where $\Delta(\Omega)$ is the Stratonovich-Weyl Kernel which is an operator-valued function on $X$. By virtue of the tracial
property, we have that
\begin{eqnarray}
{\rm Tr}[\hat A \Delta(\Omega)]&=&\int_x d\mu(\Omega^\prime) W_{A}(\Omega') W_{\Delta(\Omega)}(\Omega')\nonumber\\
&=& \int_X d\mu(\Omega^\prime) {\rm Tr} \:[\hat A \: \Delta(\Omega')] W_{\Delta(\Omega)}(\Omega')\label{tri2}
\end{eqnarray}
where $W_{\Delta(\Omega)}(\Omega')$ is the Weyl-equivalent of the Stratonovich Kernel. From (\ref{tri2}) we infer that
\be
\Delta(\Omega)= \int_X d\mu(\Omega^\prime) \Delta(\Omega')  W_{\Delta(\Omega)}(\Omega')\label{tri3}
\ee
so that the function
\be
K(\Omega, \Omega'):=W_{\Delta(\Omega)}(\Omega')={\rm Tr} [\Delta(\Omega) \Delta(\Omega')]\label{tri4}
\ee
behaves as a Dirac delta function on the manifold $X$.
Consequently, making use of this property, the Weyl rule (\ref{lin}) may be inverted to give
\be
\hat A = \int_X d\mu(\Omega) W_{A}(\Omega) \Delta(\Omega)\label{inv}
\ee
Furthermore, from (\ref{lin}), (\ref{tri3}) and (\ref{tri4}) the SW-postulates (ii)-(v) translate to the following conditions on the SW-kernel operator:
\begin{eqnarray}\nonumber
&(\rm iib)& \Delta(\Omega) = [\Delta(\Omega)]^\dag, \:\: \forall \Omega\in X\\
&(\rm iiib)& \int_X d\mu(\Omega) \Delta(\Omega)=I\nonumber\\
&(\rm ivb)& \int_X d\mu(\Omega') {\rm Tr} [\Delta(\Omega) \Delta(\Omega')]\Delta(\Omega')= \Delta(\Omega)\nonumber\\
&(\rm vb)& \Delta(g\cdot\Omega)= \pi(g) \Delta(\Omega) \pi(g)^{-1}\nonumber
\end{eqnarray}
In terms of the formalism of coherent states \cite{kg}, \cite{pere} we have that, whenever the Peter-Weyl theorem applies \cite{vari},
the SW kernel $\Delta(\Omega)$, satisfying the above conditions,
can be given explicitly as \cite {brif}:
\begin{eqnarray}
\Delta(\Omega)&=& \sum_\nu Y^\ast_\nu (\Omega) D_\nu \nonumber\\
&=& \sum_\nu  Y_\nu (\Omega) D^\dag_\nu\label{stker}
\end{eqnarray}
Here
\be
D_\nu := \int_X  d\mu(\Omega)Y_\nu (\Omega)|\Omega\rangle\langle\Omega|\label{strker2}
\ee
denotes a set of operators acting on the Hilbert space $\mathcal H$.
The harmonic functions $Y_\nu (\Omega) $, which form a complete orthonormal basis in $L^2 (X,\mu)$, are
eigenfunctions of the Laplace-Beltrami operator ($\delta d +d \delta $) associated
with the space $X$, while
the index $\nu$ is, in general, a composite label. We would like to stress here, as it should have
already become evident from our previous considerations, that since we are always going from the quantum mechanics
of operators and Hilbert space to classical phase space averages, our Weyl correspondences are surjective and therefore
unique maps (to a given quantum operator there corresponds a unique Weyl function, which corresponds to the case $s=0$ for
the families of operators and functions considered in \cite{brif}).

Note now than when substituting (\ref{stker}) and (\ref{strker2}) in (\ref{lin}) we get
\begin{eqnarray}
W_A (\Omega)&=& \sum_\nu  Y^\ast_\nu (\Omega) {\mathcal A}_\nu \nonumber\\
&=& \sum_\nu Y_\nu (\Omega) \tilde{\mathcal A}_\nu\label{weylalt}
\end{eqnarray}
where
\be
{\mathcal A}_\nu ={\rm Tr} (\hat A D_\nu)\;\;\;\;\;\;\;\;\tilde{\mathcal A}_\nu ={\rm Tr} (\hat A D^\dag_\nu)
\ee
 The generalized twisted product of two SW-symbols follows directly from (\ref{inv}) and the above and is given by
\begin{eqnarray}\label{twist}
W_A(\Omega)\star_S W_B (\Omega)&:=&W_{AB}(\Omega):= {\rm Tr} \:[\hat A \hat B\: \Delta(\Omega)]\nonumber\\
&=& \int_X d\mu(\Omega')\int_X d\mu(\Omega'') W_{A}(\Omega')W_{B}(\Omega'')L(\Omega, \Omega', \Omega'')
\end{eqnarray}
where the tri-kernel $L(\Omega, \Omega', \Omega'')$ is defined by
\be\label{trik}
L(\Omega, \Omega', \Omega''):= {\rm Tr}[\Delta(\Omega)\Delta(\Omega')\Delta(\Omega'')]
\ee

We are now ready to apply these results of the general formalism to the space-space noncommutative Heisenberg-Weyl alggroup $H_5$,
defined by the nilpotent Lie algebra (\ref{noncomm2}), for the particular case ($d=2, \; \bar\theta_{ij}=0$)
considered in the previous section. In terms of bosonic creation and destruction operators and holomorphic coordinates,
appropriate for calculating the SW kernel and symbols in terms of coherent states, the Lie algebra of the generators of $H_5$
is given by
\begin{eqnarray}\label{conm}
\left[\hat a_i, \hat a^\dag_j \right]&=&\delta_{ij}\;\;\;\;\;\;i=1,2\nonumber\\
\left[\hat a_i, \hat a_j \right]&=&\left[\hat a^\dag_i, \hat a^\dag_j \right]=0\;\;\;\;\;i=1,2
\end{eqnarray}
where
\begin{eqnarray}\label{cdop}
\hat a_1&=& (\sqrt {2\hbar })^{-1}(\hat R_1 +\frac{\theta}{2\hbar} \hat P_2 +i \hat P_1)\nonumber\\
\hat a^\dag_1&=& (\sqrt {2\hbar })^{-1}(\hat R_1 +\frac{\theta}{2\hbar} \hat P_2 -i \hat P_1)\\
\hat a_2&=& (\sqrt {2\hbar })^{-1}(\hat R_2 -\frac{\theta}{2\hbar}\hat P_1 +i \hat P_2)\nonumber\\
\hat a^\dag_2&=& (\sqrt {2\hbar })^{-1}(\hat R_2 -\frac{\theta}{2\hbar}\hat P_1 -i \hat P_2)\nonumber
\end{eqnarray}
The group elements are therefore of the form
\be\label{gre}
g(s,\alpha,\beta)= e^{(isI + \alpha \hat a_1^\dag -\bar\alpha \hat a_1 +
 \beta \hat a_2^\dag -\bar\beta \hat a_2 )}
\ee
where $\alpha, \beta\in {\Bbb C}$ and $\bar\alpha, \bar\beta$ denotes complex conjugation.
Clearly here $X= H_5/U(1)={\Bbb C}^2$, and the invariant measure is
\be\label{meas}
d\mu(\Omega) = \pi^{-2}d^2 \alpha\; d^2 \beta
\ee
The Glauber coherent states are
\be\label{gs}
|\Omega\rangle:=|\alpha,\beta\rangle = D(\alpha, \beta)|0\rangle
\ee
with $D(\alpha, \beta)$ denoting the displacement operator
\be\label{disp}
D(\alpha, \beta):=e^{(\alpha \hat a_1^\dag -\bar\alpha \hat a_1 +
 \beta \hat a_2^\dag -\bar\beta \hat a_2 )}
\ee
Since the harmonic functions in this case are the exponentials
\be\label{harmf}
Y_\nu(\Omega):= Y_{(\xi,\eta)}(\alpha,\beta)=\exp(\xi\bar\alpha -\bar\xi \alpha + \eta\bar\beta -\bar\eta \beta)
\ee
so that
\be\label{stratk2}
\Delta(\alpha,\beta)= \frac{1}{\pi^2}\int_{\Bbb C} d^2\xi \int_{\Bbb C}d^2 \eta D(\xi, \eta)
\exp(\bar\xi\alpha -\xi \bar\alpha + \bar\eta\beta -\eta \bar\beta)
\ee
the expectation value of a quantum operator $\hat A$ is given by
\begin{eqnarray}\label{expv}
\langle \hat A \rangle &=& {\rm Tr} [\hat{\boldsymbol\rho} \hat A]\nonumber\\
&=& \frac{1}{\pi^2} \int_{\Bbb C} d^2 \alpha \:\int_{\Bbb C}d^2 \beta \; W_\rho (\alpha,\beta) W_A (\alpha, \beta)
\end{eqnarray}
where
\be\label{swf}
W_\rho (\alpha,\beta) = {\rm Tr} [\Delta(\alpha,\beta)\: \hat{\boldsymbol\rho}]
\ee
is the SW-symbol corresponding to the density matrix operator $\hat{\boldsymbol\rho}$.\\
We can now make use of (\ref{twist}) and (\ref{trik}) together with (\ref{disp}) and  (\ref{stratk2}) to get
an explicit expression for the twisted product of two SW-symbols based on the quotient space ${\Bbb C}^2 = H_5/U(1)$.
Thus, noting that since the $\hat a_1 , \hat a_1^\dag$ commute with the $\hat a_2 , \hat a_2^\dag$ we can write the
displacement operator as $D(\alpha, \beta)= D(\alpha)D(\beta)$, and the tri-kernel as $L(\alpha, \alpha', \alpha''; \beta,\beta', \beta'')
=L(\alpha, \alpha', \alpha'')L( \beta,\beta', \beta'')$. Moreover,
using also repeatedly the coherent states properties
\be\label{disp2}
D(\xi)|\beta\rangle = e^{iIm(\xi\bar\beta)}|\xi+\beta \rangle
\ee
and
\be\label{inprod}
\langle \alpha|\alpha'\rangle= e^{-\frac{1}{2}(|\alpha|^2 +|\alpha'|^2 -2\bar\alpha \alpha')}
\ee
we find
\be\label{trik2}
L(\alpha, \alpha', \alpha'')= 4 \exp[4i(\alpha'{_2} \alpha_1 -\alpha'_1 \alpha_2 + \alpha'_1 \alpha''_2 - \alpha'_2 \alpha''_1 + \alpha''_1 \alpha_2 -\alpha''_2 \alpha_1)]
\ee
and an analogous expression for $L( \beta,\beta', \beta'')$.\\

Consequently
\begin{gather}
W_A(\alpha, \beta)\star_S W_B(\alpha, \beta)=\frac{16}{\pi^4}\int_{\Bbb C} d^2 \alpha''\int_{\Bbb C} d^2 \alpha'
e^{4i\alpha'_1(\alpha''_2 -\alpha_2)}e^{4i\alpha'_2(\alpha_1 -\alpha''_1)}e^{4i(\alpha''_1 \alpha_2 -\alpha''_2 \alpha_1 )}\notag\\
\qquad{}\times\int_{\Bbb C} d^2 \beta''\int_{\Bbb C} d^2 \beta'
e^{4i\beta'_1(\beta''_2 -\beta_2)}e^{4i\beta'_2(\beta_1 -\beta''_1)}e^{4i(\beta''_1 \beta_2 -\beta''_2 \beta_1 )}
W_A (\alpha',\beta')W_B (\alpha'',\beta'')\label{twist10}\hspace{1in}
\end{gather}
Making next the change of variables $\alpha''_1 =\alpha_1 +\eta_1, \; \alpha''_2 =\alpha_2 +\eta_2, \;
\beta''_1 =\beta_1 +\xi_1, \; \beta''_2 =\beta_2 +\xi_2,$ we can write
\begin{gather}\label{twist3}
W_A(\alpha, \beta)\star_S W_B(\alpha, \beta)=\frac{16}{\pi^4}\int_{\Bbb C}\dots \int_{\Bbb C}d\eta_1 d\eta_2
d\xi_1 d\xi_2 d\alpha'_1 d\alpha'_2 d\beta'_1 d\beta'_2 e^{4i(\alpha'_1 -\alpha_1)\eta_2}\notag\\
\qquad{}\times e^{-4i(\alpha'_2 -\alpha_2)\eta_1}e^{4i(\beta'_1 -\beta_1)\xi_2}e^{-4i(\beta'_2 -\beta_2)\xi_1}W_A (\alpha_1, \alpha_2, \beta_1, \beta_2)\notag\\
\qquad{}\times e^{(\eta_1 \vec{\partial}_{\alpha_1}+ \eta_2 \vec{\partial}_{\alpha_2}
+\xi_1 \vec{\partial}_{\beta_1}+ \xi_2 \vec{\partial}_{\beta_2})}W_B (\alpha_1, \alpha_2, \beta_1, \beta_2)
\end{gather}

 We can change the last exponential in the above equation into a bi-differential by noting that
\be\label{twist4}
e^{4i(\alpha'_1 -\alpha_1)\eta_2} e^{\eta_2 \vec{\partial}_{\alpha_2}}W_B (\alpha_1, \alpha_2, \beta_1, \beta_2)=
e^{4i(\alpha'_1 -\alpha_1)\eta_2} e^{-\frac{i}{4}\overleftarrow{\partial}_{\alpha'_1} \overrightarrow{\partial}_{\alpha_2}}W_B (\alpha_1, \alpha_2, \beta_1, \beta_2)
\ee
and similarly for the other terms. Hence, substituting the results in (\ref{twist10}), integrating by parts and integrating over the remaining
variables in the integrand, we finally arrive at
\be\label{twist5}
\begin{split}
W_A(\alpha, \beta)\star_S W_B(\alpha, \beta):=\hspace{3.5in}\\
W_A(\alpha,\beta)\; e^{\frac{i}{4}(\overleftarrow{\partial}_{\alpha_1} \overrightarrow{\partial}_{\alpha_2}
-\overleftarrow{\partial}_{\alpha_2} \overrightarrow{\partial}_{\alpha_1} +\overleftarrow{\partial}_{\beta_1} \overrightarrow{\partial}_{\beta_2}
-\overleftarrow{\partial}_{\beta_2} \overrightarrow{\partial}_{\beta_1})}\;W_B(\alpha,\beta) 
\end{split}
\end{equation}

Now, substituting this result into (\ref{expv}) we obtain the expectation value of a product of quantum operators
derived according to the Stratonovich-Weyl correspondence in the context of the space-space noncommutative Heisenberg-Weyl group. Moreover,
since the alternate calculation in the previous section was done based on the Lie algebra of the same group, and
since the Stratonovich phase-space formulation was purported to be a generalization of the later to physical systems with
Lie group symmetries which, evidently include the one common to the two approaches, a coincidence of results would then appear natural. In order to verify this conjecture we first need to convert the holomorphic variables in (\ref{expv}), (\ref{swf}) and (\ref{twist5}) into phase-space variables. That is, we need to make the substitutions:
\begin{eqnarray}\label{chvar1}
\alpha_1 \to \frac{1}{\sqrt{2\hbar}}(q_1 +\frac{\theta}{2\hbar}p_2)&\;\;\;&\alpha_2 \to \frac{1}{\sqrt{2\hbar}}p_1 \nonumber\\
 \beta_1 \to \frac{1}{\sqrt{2\hbar}}(q_2 -\frac{\theta}{2\hbar}p_1)&\;\;\;&\beta_2\to\frac{1}{\sqrt{2\hbar}}p_2
\end{eqnarray}
Hence,
\begin{eqnarray}\label{chvar}
\partial_{\alpha_1}&=&\sqrt {2\hbar}\partial_{q_1} \;\;\; \;\;\;\;\;\;\;\partial_{\alpha_2}=\sqrt {2\hbar}
(\frac{\theta}{2\hbar}\partial_{q_2}+ \partial_{p_1})\nonumber\\
\partial_{\beta_1}&=& \sqrt {2\hbar}\partial_{q_2} \;\;\;\;\;\;\;\;\;\; \partial_{\beta_2}=\sqrt {2\hbar}
(-\frac{\theta}{2\hbar}\partial_{q_1}+ \partial_{p_2})
\end{eqnarray}
from where the Stratonovich twist bi-differential expressed in terms of phase-space variables
takes the form
\be\label{bigstar}
\star_S = \star_\theta \circ \star_\hbar
\ee
Furthermore, making use of (\ref{disp}), (\ref{harmf}), (\ref{swf}) and (\ref{stratk2}), we have
\be\label{swf2}
\begin{split}
W_\rho (\alpha,\beta) = {\rm Tr} [\Delta(\alpha,\beta)\: \hat{\boldsymbol\rho}]=\hspace{2in}\nonumber\\
 \frac{1}{\pi^2}\int_{\Bbb C} d^2\xi \int_{\Bbb C}d^2 \eta
{\rm Tr}[ e^{(\xi\hat a_1^\dag -\bar\xi \hat a_1 +
 \eta \hat a_2^\dag -\bar\eta \hat a_2 )} \hat{\boldsymbol\rho}]
\exp(\bar\xi\alpha -\xi \bar\alpha + \bar\eta\beta -\eta \bar\beta)
\end{split}
\ee
Evaluating now the trace in the above expression relative to the mixed phase-space basis $\{|q_1, p_2 \rangle \}$
and after a fairly lengthy but straightforward calculation we arrive at
\be\label{swf3}
\begin{split}
W_\rho (\alpha,\beta)= 4\int\int dq'_1 dp'_2 e^{2i\alpha_2 (2\alpha_1 -{\sqrt\frac{2}{\hbar}}q'_1-
\frac{\theta}{\hbar\sqrt{2\hbar}}p'_2)} e^{-2i\beta_1(2\beta_2 -{\sqrt\frac{2}{\hbar}}p'_2 )}\\
\times \langle q'_1, p'_2|\hat{\boldsymbol\rho}|2\sqrt{2\hbar}\alpha_1 -q'_1 - \frac{2\theta}{\sqrt{2\hbar}}\beta_2,
-p'_2 + 2\sqrt{2\hbar}\beta_2 \rangle
\end{split}
\ee
Finally, making the change of variables
\begin{eqnarray}\label{chvar}
q'_1 &=& \sqrt{2\hbar}\alpha_1 -\frac{\lambda_1}{2}-\frac{\theta}{\sqrt{2\hbar}}\beta_2\nonumber \\
p'_2&=& \beta_2 -\frac{\lambda_2}{2}
\end{eqnarray}
yields
\be\label{swf4}
\begin{split}
W_{\rho} (\alpha,\beta)=\int\int d\lambda_1 d\lambda_2 e^{\frac{2i\alpha_2}{\sqrt{2\hbar}}
(\lambda_1 + \frac{\theta}{2\hbar}\lambda_2)} e^{-\frac{2i\beta_1\lambda_2}{\sqrt{2\hbar}}}\hspace{1.5in}\\
\times \langle \sqrt{2\hbar}\alpha_1 -\frac{\lambda_1}{2}-\frac{\theta}{\sqrt{2\hbar}}\beta_2 ,\beta_2 -\frac{\lambda_2}{2}|
\hat{\boldsymbol\rho}|\sqrt{2\hbar}\alpha_1 +\frac{\lambda_1}{2}-\frac{\theta}{\sqrt{2\hbar}}\beta_2 ,\beta_2 +
\frac{\lambda_2}{2}\rangle
\end{split}
\ee
In terms of phase-space variables this result reads
\be\label{swf5}
\begin{split}
W_{\rho} (\alpha(p_1,q_2),\beta(q_1,p_2))=e^{-\frac{\theta}{\hbar}p_1\partial_{q_2}}
\int\int d\lambda_1 d\lambda_2 e^{\frac{i}{\hbar}(p_1 \lambda_1 - q_2 \lambda_2)}\\
\times \langle q_1 -\frac{\lambda_1}{2}, p_2 -\frac{\lambda_2}{2}|\hat{\boldsymbol\rho}|
q_1 +\frac{\lambda_1}{2}, p_2 +\frac{\lambda_2}{2}\rangle
\end{split}
\ee

If we now compare Eqs. (\ref{twist5}), (\ref{bigstar}) and (\ref{swf5}) with (\ref{starproduct3}), (\ref{starproduct4}), (\ref{2.31})
and (\ref{weylwig3}) of the previous section, we see that for the space-space noncommutative Weyl-Heisenberg Lie group the quantum mechanics' resulting
from both formalisms are equivalent provided that in the calculation of the expectation values we derive the phase-space averages by
combining the appropriate $\star$-product for the evaluation of Weyl-symbols with the appropriate Wigner function or Weyl-symbol associated
with the density matrix for the problem, according to the above referred formulas.

\section{The Berezin quantization procedure by means of involution operators and its application to the space-space noncommutative Heisenberg-Weyl algebra}

This quantization scheme arises from the basic property that for homogeneous symmetric spaces there is an involutive automorphism of $G$ acting on them. Such is the case
for $X= H_5/U(1)$ where the involution automorphisms are reflections around each point. Recalling equations (\ref{op}, \ref{alpha}) in Sec.2, we see that the Weyl function is the Fourier transform
of the $\alpha$ function in (\ref{op}) while the Fourier transform of the unitary displacement operators $\{(2\pi\hbar)^{-1} \exp[\frac{i}{\hbar}({\bf y}\cdot {\bf\hat  R}
+{\bf x}\cdot{\bf\hat  P})]\}$ are indeed reflections. It is thus natural to write \cite{moreno, bere}
\be\label{weylinvolution}\hat A=\int_X d\mu (x) w_A(x) \hat U(x)\ee
as a generalization of ((\ref{op}).
Here $\hat U(x)$ is the unitary operator corresponding to the group element that performs reflections around the point $x\in X$.\\
As noted by the authors in \cite{moreno}, the use of the reflection operator provides a way to circumvent the situation when a Fourier transform on $X$ cannot be consistently defined. The function $w_A(x)$ appearing in (\ref{weylinvolution}) corresponds to the Weyl contravariant symbol which is, in general, different from the Weyl covariant symbol defined as:
\be\label{weylcov} {\tilde w_A(x)}:=\mathrm{Tr}[{\hat A}{\hat U(x)}]\ee
Berezin also showed that there exists a bijective map relating $w_A, {\tilde w_A}$ to the usual contravariant and covariant symbols $P_A, Q_A$ respectively, whose expressions are given by
\begin{gather}\label{beresymbols} {\hat A}=\int_X d\mu(x)P_A(x)|x\rangle\langle x|\\
Q_A(x)=\langle x|{\hat A}|x\rangle\end{gather}
where $\{|x\rangle\}$ corresponds to an overcomplete basis of normalized states tagged by points in $X$.\\
Thus in order to implement this quantization formalism we must first determine what will be in our case the reflection operator $\hat U(x)$. To this end we will make use of the Hilbert space spanned by the coherent states of the last section, which in fact constitute an overcomplete basis. Each coherent state $|\alpha,\beta\rangle=|\alpha\rangle\otimes|\beta\rangle$ is tagged by a point $(\alpha,\beta)\in\mathbb{C}^2=X$.\\
We may now construct the reflection operator $\hat U(\alpha,\beta)$ by acting transitively on the reflection operator around the origin $\hat U(0,0)$ with the unitary operator associated to $g\in G$. From the properties of the algebra (\ref{conm}) it is clear that ${\hat U(\alpha,\beta)}={\hat U(\alpha)}\otimes{\hat U(\beta)}$, where each $\hat U(\alpha)$ acts on a copy of $\mathbb{C}$. Then for simplicity we will reduce the calculation to one copy of $\mathbb{C}$ and obtain the final result just by taking the direct product of the two copies. Thus, following Berezin, consider a complex line bundle $L$ over $\mathbb{C}$ with fiber metric $e^{-K(v,\bar v)}$, where $K(v,\bar v)=v\bar v$ is the K\"ahler potential. The Hilbert space $\mathcal{H}$ consists of holomorphic sections of $L$ with inner product
\be\label{innerberg} \langle f|g\rangle={1\over\pi}\int_\mathbb{C} d^2v\;{\bar f(v)}g(v)e^{-v\bar v}\ee
where the holomorphic section $f(v)$ denotes the evaluation
\be f(v)=\langle v|f\rangle\ee
The coherent state $|\alpha\rangle$, expressed in the Fock-Bargmann representation $\frak{F}$, is given by
\be |\alpha\rangle=e^{-{1\over 2}|\alpha|^2}\sum_{n=0}^{\infty} {\alpha^n\over\sqrt{n!}}|n\rangle\ee
Hence
\be\label{ev2} \langle v|\alpha\rangle=e^{-{1\over2}|\alpha|^2}\sum_{n=0}^{\infty}\frac{\alpha^{n} \bar{v}^{n}}{n!}=e^{-{1\over 2}|\alpha|^2+\alpha\bar v}\ee
Making use of the identity resolution
\be\label{ires} \mathbb{I}={1\over\pi}\int_\mathbb{C} d^2v\;e^{-|v|^2}|v\rangle \langle v|    \ee
 we can write the left hand of (\ref{ev2}) as
\be\label{ev3} \alpha(v)={1\over\pi}\int_{\mathbb{C}}d^2v'\langle v|v'\rangle e^{-|v'|^2}\alpha(v')\ee
It is easy to show that this equation becomes an identity if we set $\langle v|v'\rangle:= B(v',{\bar v})= e^{v'v}$ and make use of (\ref{ev2}) on both sides of the equation.                     
Moreover, it also follows that $ B(v',\bar v)$ satisfies the following properties
\be{1\over\pi}\int_\mathbb{C} d^2v'\;e^{-|v'|^2}B(v',{\bar v})f(v')=f(v)\nonumber\ee
\be\label{bergman} {1\over\pi}\int_\mathbb{C} d^2v'\;e^{-|v'|^2}B(v,{\bar v'})B(v',{\bar u})=B(v,{\bar u})\ee
Thus $B(v',{\bar v})$   is the Bergman reproducing kernel \cite{berg}, and in the $\frak F$ representation space the quantity $\pi\delta (v,v'):= B(v',\bar v)e^{-|v'|^2}$ acts as a Dirac delta function under integration.\\

Let us now define the operator $\hat U(0)$ by
\be\label{inv0} {\hat U(0)}:={1\over\pi}\int_\mathbb{C}d^2v\;e^{-|v|^2}|-v\rangle\langle v|\ee
To show that this is the reflection operator around the origin we take the action of $\hat U(0)$ over any arbitrary state $|v'\rangle$ and use the above definition of the delta function action:
\begin{gather}\label{reflex}{\hat U(0)}|v'\rangle={1\over\pi}\int_\mathbb{C}d^2 v\; e^{-|v|^2}\;|-v\rangle\langle v|v'\rangle = {1\over\pi}\int_\mathbb{C} d^2v\;e^{-|v|^2}B(v',{\bar v})|-v\rangle =|-v'\rangle\end{gather}
With the above results, we are now in a position to calculate the more general operator $\hat U(\zeta)$. This is done by noticing that by taking the unitary transformation ${\hat D(\zeta)}{\hat U(0)}{\hat D^{\dag}(\zeta)}$, where $\hat D(\zeta)$ is the unitary displacement operator representation of the $H_3$ group acting on coherent states according to (\ref{disp2}). Since $\hat U(0)$ is an involution,
$\hat D(\zeta)$ induces displacements and $({\hat D(\zeta)}{\hat U(0)}{\hat D^{\dag}(\zeta)
})^2=\mathbb{I}$, the operator $\hat U(\zeta)$ must correspond to a reflection around $\zeta\in\mathbb{C}$.
To show this we use first Eq.(\ref{disp}) to obtain the explicit form of the operator ${\hat U(\zeta)}:={\hat D(\zeta)}{\hat U(0)}{\hat D^{\dag}(\zeta)}$:
\be\label{inv1} {\hat U(\zeta)}={1\over\pi}\int_\mathbb{C} d^{2}v  e^{-|v|^2}{\hat D(\zeta)} \; |-v\rangle\langle v|\hat D^{\dag}(\zeta)\ee
Making now use of (\ref{ev2}) in order to express the arbitrary ket $|v\rangle$ in terms of the normalized coherent state basis, {\it i.e.} as
\be |v\rangle={1\over\pi}\int_\mathbb{C} d^2\alpha\; e^{(-\frac{1}{2}|\alpha|^2 + \bar{\alpha}v)} |\alpha\rangle\ee
 and applying (\ref{disp2}) on the coherent state $|\alpha\rangle$ yields
\be\label{gopact} {\hat D(\zeta)}|v\rangle = {1\over\pi}\int_\mathbb{C} d^2\alpha\; e^{(-\frac{1}{2}|\alpha|^2 + \bar{\alpha}v +i\text {Im}(\zeta\bar\alpha))} |\alpha +\zeta\rangle \ee
Furthermore making use of (\ref{gopact}) and the properties of the Bergman kernel in (\ref{bergman}) we obtain after some fairly straightforward calculations the expression
\be\label{inv2} {\hat U(\zeta)}={1\over\pi}\int_\mathbb{C}d^2\alpha\; e^{(\zeta\bar\alpha-\bar\zeta\alpha)}|\alpha+\zeta\rangle\langle\zeta-\alpha|\ee
Finally, making the change of variables $\zeta-\alpha =\rho$ yields
\be\label{inv3} {\hat U(\zeta)}={1\over\pi}\int_\mathbb{C}d^2\rho\;e^{{\bar\zeta}\rho-{\bar\rho}\zeta}|2\zeta-\rho\rangle\langle\rho|\ee
We next use this expression to repeat a similar calculation to the one we did above in order to obtain $\hat U(0)$. Thus, taking the action of the operator $\hat U(\zeta)$ on an arbitrary state $|v\rangle$, and expanding the coherent state $|2\zeta-\rho\rangle$ in (\ref{inv3}) in terms of $|v\rangle$, by making use of (\ref{ev2}) and (\ref{ires}), we get
\be{\hat U(\zeta)}|v\rangle={1\over\pi^2}e^{-2|\zeta|^2}\int_{\mathbb{C}}d^2v'\;e^{-|v'|^2}e^{2{\bar v'}\zeta}|v'\rangle\int_{\mathbb{C}}d^2\rho\;e^{-|\rho|^2}e^{v\bar\rho}e^{(2{\bar\zeta}-{\bar v'})\rho}\ee
which when resorting repeatedly to equation (\ref{bergman})  gives
\be\label{inv4} {\hat U(\zeta)}|v\rangle=e^{2({\bar\zeta}v-|\zeta|^2)}|2\zeta-v\rangle\ee
The function inside the ket in the above equation can be rewritten as $2(\zeta-v)+v$ to make evident the fact that this is the reflection of the point $v$ around $\zeta$. To complete the proof we check that $\hat U(\zeta)$ is indeed an involution. This follows directly by  once more acting with $\hat U(\zeta)$ on Eq.(\ref{inv4}). Accordingly we obtain
\begin{align} {\hat U(\zeta)}^2|v\rangle&={\hat U(\zeta)}[e^{2({\bar\zeta}v-|\zeta|^2)}|2\zeta-v\rangle]\nonumber\\
&=e^{2({\bar\zeta}v-|\zeta|^2)}e^{2{\bar\zeta}(2\zeta-v)}e^{-2|\zeta|^2}|2\zeta-(2\zeta-v)\rangle =|v\rangle.\end{align}

As we mentioned at the beginning of this section the Weyl contravariant and covariant symbols are  not the same in general. We will show, however, that for the symmetric homogeneous space treated here this is not the case. Indeed, making the change ${\hat U(\zeta)}\rightarrow 2{\hat U(\zeta)}\equiv\hat V(\zeta)$ in equation (\ref{weylinvolution}) the latter reduces to equation (\ref{inv}) and consequently $w_A=W_A=\tilde w_A$ in which case both symbols are equal.
This follows from equation (\ref{inv3}) and observing that by using our previous results we can write the identity
\be\label{trket}e^{{\bar\zeta}\rho-{\bar\rho}\zeta}|2\zeta-\rho\rangle={1\over2\pi}\int_\mathbb{C}d^2\lambda\;e^{{\bar\lambda}\zeta-{\bar\zeta}\lambda}
e^{{1\over2}({\bar\rho}\lambda-{\bar\lambda}\rho)}|\lambda+\rho\rangle\ee
Moreover, the coherent state $e^{{1\over2}({\bar\rho}\lambda-{\bar\lambda}\rho)}|\lambda+\rho\rangle$ is nothing else but ${\hat D(\lambda)}|\rho\rangle$ so we can replace this into (\ref{inv3}) and the operator $\hat V(\zeta)=2{\hat U(\zeta)}$ takes now the form
\be\label{inv5}{\hat V(\zeta)}={1\over\pi^2}\int_\mathbb{C}\int_\mathbb{C}d^2\lambda d^2\rho\;e^{{\bar\lambda}\zeta-{\bar\zeta}\lambda} {\hat D(\lambda)}|\rho\rangle\langle\rho|\ee
Finally observe that in this last expression the quantity ${1\over\pi}\int_\mathbb{C}d^2\rho\;|\rho\rangle\langle\rho|$ is just the identity operator in terms of normalized coherent states. It is
then obvious that equation (\ref{inv5}) reduces simply to
\be\label{inv6}{\hat V(\zeta)}={1\over\pi}\int_\mathbb{C}d^2\lambda\;e^{{\bar\lambda}\zeta-{\bar\zeta}\lambda}{\hat D(\lambda)}\ee
which allows us to conclude that ${\hat V(\alpha)}\otimes{\hat V(\beta)}\equiv\Delta(\alpha,\beta)$ as seen from equation (\ref{stratk2}). This argument demonstrates that for the Heisenberg-Weyl algebra (\ref{noncomm2}) the SW formalism as well as that of Berezin provide the same quantization scheme.\\

It is interesting to observe that, because both the SW and the Berezin formalisms are based on complex valued holomorphic states and non-Hermitian operators, defined in turn by means of creation and destruction operators, the noncommutativity of the observables in the algebra (\ref{noncomm2})is hidden in the definition of those creation and destruction operators. So, as long as we remain in the complex domain, their quantum mechanics' for the ordinary and the Heisenberg-Weyl algebras (\ref{noncomm2}) appear as indistinguishable (see {\it e.g.} equation (\ref{twist10})). It should also be clear from our presentation so far that there are a variety of Bopp maps that can be chosen to construct creation and destruction operators from phase-space operator observables. In our construction (see (\ref{cdop})) we have chosen a map that keeps the algebra of $\hat a$ and $\hat a^\dag$ unchanged, as this choice allows us to use all the machinery of standard WWGM up to the point where we re-express the final results in terms of real dynamical phase-space variables. \\
Moreover, it is known that for the WWGM quantum mechanics there is a $\star$-value equation which is a result stronger than the one providing the phase-space
expectation values for operators and products of operators on Hilbert space. Indeed, it is fairly straightforward to show that ({\it c.f. e.g.} \cite{Sz})
the star-value equation
\be\label{starv1}
W_H ({\bf p}, {\bf q}) \star_{\hbar} \rho_w =E \rho_w 
\ee
is a necessary and sufficient condition for the weaker expectation value relation
\be\label{starv2}
\int\int d{\bf p} d{\bf q} W_H ({\bf p}, {\bf q}) \: \rho_w =\int\int d{\bf p} d{\bf q} W_H ({\bf p}, {\bf q}) \star_\hbar \rho_w 
\ee
to follow. Here $W_H ({\bf p}, {\bf q}) $  is the Weyl-symbol associated with the Hamiltonian operator $\hat H$ satisfying the eigenvalue equation
$\hat H|\Psi\rangle = E |\psi\rangle$, $|\Psi\rangle$ is a pure energy state and $\rho_w$ is the Wigner function corresponding to the pure
state density matrix $\hat {\boldsymbol \rho} =|\psi\rangle\langle\psi|$. We shall investigate next if similar $\star$-valued equations exist for the quantum mechanical formulations
on the Weyl-Heisenberg group consider above, and whether their equivalence stands for such stronger equations.

\section{Star-value equations for phase-space quantum mechanics based on the space-space noncommutative Heisenberg-Weyl group }

Given a Hamiltonian $\hat H({\bf \hat P, \bf \hat R})$ for a quantum mechanical system where $\bf \hat P, \bf \hat R$  satisfy
the algebra (\ref{noncomm2}) (with $i,j=1,2$ and $\bar\theta=0$) and the pure state density matrix $\hat {\boldsymbol \rho} =|\psi\rangle\langle\psi|$,
we can consider star-value equations associated with the $\star$-products (\ref{starproduct2}) or (\ref{starproduct4}). Let us begin
by considering first the $\star$-product in  (\ref{starproduct2}) between the Weyl-symbol corresponding to $\hat H$ and the Weyl-symbol
corresponding to the density matrix $\hat {\boldsymbol \rho} $. We get (after resorting to (\ref{weylwig3}) in order to obtain the last equality):
\begin{eqnarray}\label{ncstarv1}
W_H \star W_\rho &=&
m\circ \left[e^{\sum_{i=1,2}\frac{i\hbar}{2}(\partial_{q_i}\otimes\partial_{p'_i}-\partial_{q'_i}\otimes\partial_{p_i})}
\circ
e^{\frac{\theta}{\hbar}p_1\partial_{q_2}}\otimes e^{\frac{\theta}{\hbar}p'_1
\partial_{q'_2}}W_{H}({\bf p},{\bf q})\otimes W_{\rho}({\bf p'},{\bf q'})\right]_{\bf{q,p=q',p'}}\nonumber\\
&=& (e^{\frac{\theta}{\hbar}p_1 \partial_{q_2}} W_H)\star_\hbar (e^{\frac{\theta}{\hbar}p_1 \partial_{q_2}} W_\rho) =
(e^{\frac{\theta}{\hbar}p_1 \partial_{q_2}} W_H)\star_\hbar \rho_w. 
\end{eqnarray}

Note that in general $$e^{\frac{\theta}{\hbar}p_1 \partial_{q_2}} W_H ({\bf p}, {\bf q})= W_H \left({\bf p}, q_1, q_2 +\frac{\theta}{\hbar}p_1\right) $$
which says: calculate first the Weyl-symbol corresponding to the Hamiltonian operator by applying (\ref{starproduct2})  repeatedly, followed by the displacement
of the $q_2$ argument by the exponential on the left hand side of the above expression. Hence
\be\label{ncstarv2}
W_H \star W_\rho = W_H \left({\bf p}, q_1, q_2 +\frac{\theta}{\hbar}p_1\right)\star_\hbar \rho_w
\ee
Substituting now the expression (\ref{2.31}) for the Wigner function and (\ref{qstar}) for the $\star_\hbar$-product we have
\be\label{ncstarv3}
\begin{split}
W_H \star W_\rho = (2\pi\hbar)^{-2}\int\int ds_1 ds_2 \psi(q_1 - \frac{s_1}{2}, p_2 -\frac{s_2}{2}) \psi^\ast(q_1 + \frac{s_1}{2},p_2 +\frac{s_2}{2} ) \hspace{1in}\\
\times\left[\hat W_H \left(q_1, q_2 + \frac{i\hbar}{2} \overrightarrow{\partial}_{p_2}+
\frac{\theta}{\hbar} (p_1 -\frac{i\hbar}{2} \overrightarrow{\partial}_{q_1} ); p_1 -\frac{i\hbar}{2} \overrightarrow{\partial}_{q_1}, p_2 \right)
e^{\frac{i}{\hbar}s_1(p_1 +\frac{i\hbar}{2}\overleftarrow{\partial}_{q_1})}e^{-\frac{i}{\hbar}s_2(q_2 -\frac{i\hbar}{2}\overleftarrow{\partial}_{p_2})}\right]\\
=(2\pi\hbar)^{-2}\int\int ds_1 ds_2 \psi(q_1 - \frac{s_1}{2}, p_2 -\frac{s_2}{2}) \psi^\ast(q_1 + \frac{s_1}{2},p_2 +\frac{s_2}{2} )\hspace{1.7in} \\
\times\left[\hat W_H \left(q_1 -\frac{s_1}{2}, q_2 + \frac{i\hbar}{2} \overrightarrow{\partial}_{p_2}+
\frac{\theta}{\hbar} p_1 -\frac{i\theta}{2} \overrightarrow{\partial}_{q_1} ; p_1 -\frac{i\hbar}{2} \overrightarrow{\partial}_{q_1}, p_2 -\frac{s_2}{2}\right)
e^{\frac{i}{\hbar}s_1  p_1} e^{-\frac{i}{\hbar}s_2 q_2 }\right]
\end{split}
\ee
If we now note that we can make the following replacement of the $q_2$ and $p_1$ arguments in $W_H$ inside the square brackets:
$$ q_2 \to i\hbar \partial_{s_2}\;\;\;\;\;\;\; p_1 \to -i\hbar\partial_{s_1} $$
and integrate by parts, we arrive at
\be\label{ncstarv4}
\begin{split}
W_H \star W_\rho = (2\pi\hbar)^{-2}\int\int ds_1 ds_2 e^{\frac{i}{\hbar}s_1  p_1} e^{-\frac{i}{\hbar}s_2 q_2 }\hspace{2in}\\
\times\left[\hat W_H \left(q_1 -\frac{s_1}{2}, -i\hbar\partial_{s_2}+ \frac{i\hbar}{2} \overrightarrow{\partial}_{p_2} + i\theta\partial_{s_1} -\frac{i\theta}{2} \overrightarrow{\partial}_{q_1};\;
i\hbar \partial_{s_1} -\frac{i\hbar}{2} \overrightarrow{\partial}_{q_1} ;  p_2 -\frac{s_2}{2}\right)\right. \\
\left.\psi(q_1 - \frac{s_1}{2}, p_2 -\frac{s_2}{2})
\psi^\ast(q_1 + \frac{s_1}{2},p_2 +\frac{s_2}{2} )\right]
\end{split}
\ee
Observe next that making the identifications
\begin{eqnarray}\label{ncstarv5}
\hat Q_1 :&=& q_1 -\frac{s_1}{2}\;\;\;\;\;\;\;\;\;\;\;
 \hat\Pi_1 :=i\hbar \partial_{s_1} -\frac{i\hbar}{2} {\partial}_{q_1} \nonumber\\
\hat\Pi_2 :&=&p_2 -\frac{s_2}{2}\;\;\;\;\;\;\;\;\;\;\;\hat Q_2 :=-i\hbar\partial_{s_2}+ \frac{i\hbar}{2}{\partial}_{p_2} +\frac{\theta}{\hbar}\hat\Pi_1
\end{eqnarray}
we obtain a realization for the Heisenberg-Weyl algebra:
\be
[\hat Q_1, \hat Q_2]=i\theta\,\,\,\,\;\;\;\; [\hat Q_i, \hat \Pi_j] =i\hbar \delta_{ij}\,\,\,\;\;\;\;
[\hat \Pi_1, \hat \Pi_2]= 0.
\ee
Observe also that the operator $\hat W_H (\hat Q_1, \hat Q_2, \hat\Pi_1, \hat\Pi_2)$ annihilates any function of $q_1 +\frac{s_1}{2}$ and $p_2 +\frac{s_2}{2}$.
Hence
\be\label{ncstarv6}
\begin{split}
W_H \star W_\rho = (2\pi\hbar)^{-2}\int\int ds_1 ds_2 e^{\frac{i}{\hbar}s_1  p_1} e^{-\frac{i}{\hbar}s_2 q_2 }\psi^\ast(q_1 + \frac{s_1}{2},p_2 +\frac{s_2}{2} )\\
\times\left[\hat W_H (\hat Q_1, \hat Q_2, \hat \Pi_1, \hat \Pi_2) \psi(q_1 - \frac{s_1}{2}, p_2 -\frac{s_2}{2})
\right]
\end{split}
\ee
Furthermore, consider the eigenvalue equation
\be\label{eigen}
\hat H(\hat R_1, \hat R_2; \hat P_1, \hat P_2 )|\psi\rangle = E|\psi\rangle
\ee
Since the operators $\bf \hat P, \bf \hat R$  satisfy
the algebra (\ref{noncomm2}) (with $i,j=1,2$ and $\bar\theta=0$), the projection of (\ref{eigen}) with the bra $\langle R_1, P_2|$
yields (making use of (\ref{2.2})):
\be\label{eigen2}
\hat H (R_1, -i\theta\partial_{R_1} +i\hbar\partial_{P_2}; -i\hbar \partial_{R_1}, P_2)\langle  R_1, P_2|\psi\rangle= E\langle  R_1, P_2|\psi\rangle
\ee
Setting now
\be\label{eigen3}
R_1 \equiv {\hat Q_1} = q_1 -\frac{s_1}{2}\;\;\;\;\;\;\;\; P_2 \equiv {\hat\Pi_2}= p_2 -\frac{s_2}{2}
\ee
and comparing the expression for $\hat R_2 =-i\theta\partial_{R_1} +i\hbar\partial_{P_2}$ in (\ref{eigen2}) with $\hat Q_2$ in (\ref{ncstarv5}) we get
\be\label{eigen5}
\partial_{R_1}=\frac{1}{2}\partial_{q_1}-\partial_{s_1}\qquad
\partial_{P_2}=\frac{1}{2}\partial_{p_2}-\partial_{s_2}
\ee
On the other hand also comparing the $\hat R_2$ in (\ref{eigen2}) with (\ref{2.2}) yields
\be\label{eigen4}
\partial_{q_1}= \partial_{R_1}\qquad\partial_{p_2}= \partial_{P_2}
\ee
from where it also clearly follows
\be\partial_{s_1} = -\frac{1}{2}\partial_{R_1}\qquad\partial_{s_2} = -\frac{1}{2}\partial_{P_2}\ee

Substituting the above into (\ref{eigen2}) and comparing with (\ref{ncstarv5}) we arrive at
\be\label{eigen6}
\hat H (\hat Q_1, \hat Q_2; \hat \Pi_1, \hat \Pi_2)\langle  Q_1, \Pi_2|\psi\rangle= E\langle  Q_1, \Pi_2|\psi\rangle
\ee
so, if we could make the identification $\hat H (\hat Q_1, \hat Q_2; \hat \Pi_1, \hat \Pi_2)=W_H (\hat Q_1, \hat Q_2, \hat \Pi_1, \hat \Pi_2)$,
we would then have that (\ref{ncstarv6}) would immediately imply that
\begin{eqnarray}\label{ncstarv7}
  W_H \star W_\rho  &=& (2\pi\hbar)^{-2}E\int\int ds_1 ds_2 e^{\frac{i}{\hbar}s_1  p_1} e^{-\frac{i}{\hbar}s_2 q_2 }
\psi^\ast(q_1 + \frac{s_1}{2},p_2 +\frac{s_2}{2} )\psi(q_1 - \frac{s_1}{2}, p_2 -\frac{s_2}{2})\nonumber \\
&=& E\rho_w 
\end{eqnarray}
or
\be\label{eigen7}
W_H ({\bf p}; q_1, q_2 + \frac{\theta}{\hbar}p_1)\star_{\hbar}\rho_w({\bf p}, {\bf q})= E\rho_w
\ee
Note, however that the feasibility of this identification requires that $\hat H (\hat Q_1, \hat Q_2; \hat \Pi_1, \hat \Pi_2)$ and
$W_H (\hat Q_1, \hat Q_2, \hat \Pi_1, \hat \Pi_2)$ should be of the same functional form for their operator arguments. But, according to our discussion
following equation (\ref{weylprod}) this will only be possible for Hamiltonians having the Weyl symmetrized ordering of operators.\\
The corresponding expression of the $\star$-value equation for the product $W_H \star_\theta\circ\star_\hbar W_\rho$ follows immediately by recalling (see the argument given in the paragraph following equation (\ref{inv6})) that in holomorphic coordinates the $\star$-value equation does not see the non-commutativity, {\it i.e.}
\be W_H(\alpha,\beta)\star_SW_\rho(\alpha,\beta)=EW_\rho\equiv W_H(\alpha,\beta)\star_\hbar W_\rho(\alpha,\beta)=EW_\rho\ee
Thus when going back to phase-space variables by making use of (\ref{chvar1}) and (\ref{swf5}) yields
\be\label{eigen8}\begin{split} W_H({1\over\sqrt{2\hbar}}(q_1+{\theta\over\sqrt{2\hbar}}p_2),{1\over\sqrt{2\hbar}}(q_2-{\theta\over\sqrt{2\hbar}}p_1),{1\over\sqrt{2\hbar}}p_1,{1\over\sqrt{2\hbar}}p_2)\star_\theta\circ\star_\hbar e^{-{\theta\over\hbar}p_1\partial_{q_2}}\rho_w(q_1,q_2,p_1,p_2)=\\
E e^{-{\theta\over\hbar}p_1\partial_{q_2}}\rho_w(q_1,q_2,p_1,p_2)\end{split}\ee
Evidently the two $\star$-valued equations (\ref{eigen7}) and (\ref{eigen8}) are different, even that the weaker expectation values resulting from them are the same. This difference may turn out to be important for certain problems in deformation quantization such as the ones mentioned in the introduction.

\section{Acknowledgement}
This work was supported in part by CONACyT Project  UA7899-F.

\def\BibTeX{{\rm B\kern-.05em{\sc i\kern-.025em b}\kern-.08em
     T\kern-.1667em\lower.7ex\hbox{E}\kern-.125emX}}


\label{lastpage}

\end{document}